\documentclass[sn-aps]{sn-jnl}% American Physical Society (APS) Reference Style
\usepackage{graphicx}%
\usepackage{multirow}%
\usepackage{amsmath,amssymb,amsfonts}%
\usepackage{amsthm}%
\usepackage{mathrsfs}%
\usepackage[title]{appendix}%
\usepackage{xcolor}%
\usepackage{textcomp}%
\usepackage{manyfoot}%
\usepackage{booktabs}%
\usepackage{algorithm}%
\usepackage{algorithmicx}%
\usepackage{algpseudocode}%
\usepackage{listings}%
\usepackage{bm}
\usepackage{bbm}
\theoremstyle{thmstylethree}%

\begin{document}

\title[]{Confinement of $N$-body systems and non-integer dimensions}


\author*[1]{\fnm{E.} \sur{Garrido}}\email{e.garrido@csic.es}

\author[2]{\fnm{A.S.} \sur{Jensen}}\email{asj@phys.au.dk}
\equalcont{These authors contributed equally to this work.}


\affil*[1]{\orgdiv{Instituto de Estructura de la Materia}, \orgname{IEM-CSIC}, \orgaddress{ \city{Madrid}, \postcode{E-28006}, \country{Spain}}}

\affil[2]{\orgdiv{Department of Physics and Astronomy}, \orgname{Aarhus University}, \orgaddress{ \city{Aarhus C}, \postcode{DK-8000}, \country{Denmark}}}

%\affil[3]{\orgdiv{Department}, \orgname{Organization}, \orgaddress{\street{Street}, \city{City}, \postcode{610101}, \state{State}, \country{Country}}}


\abstract{The squeezing process of a three-dimensional quantum system by use of an external
deformed one-body oscillator potential can also be described by the $d$-method,
without external field and where the dimension can take non-integer values.
In this work we first generalize both methods to $N$ particles and any transition 
between dimensions below $3$. Once this is done, the use of harmonic oscillator interactions 
between the particles allows complete analytic solutions of both methods, and a direct comparison 
between them is possible.
Assuming that both methods describe the same process, leading to the same 
ground state energy and wave function, an analytic equivalence between the methods arises.

The equivalence between both methods and the validity of the derived analytic
relation between them is first tested for two identical bosons and for squeezing transitions
from 3 to 2 and 1 dimensions, as well as from 2 to 1 dimension. We also investigate the symmetric 
squeezing from 3 to 1 dimensions of a system made of three identical bosons. We have in all the cases found that the derived analytic relations between the two methods work
very well. This fact permits to relate both methods also for large
squeezing scenarios, where the brute force numerical calculation with the external field
is too much demanding from the numerical point of view, especially for systems with more than two particles.
}

%\keywords{keyword1, Keyword2, Keyword3, Keyword4}



\maketitle

\section{Introduction}\label{sec1}

A system may be constrained to a phase-space smaller than allowed in
our ordinary three-dimensional world \cite{pet00,gre01,blo05,nis08}.  The corresponding effective
dimension does not have to be an integer like 3, 2, 1, or 0
\cite{gar19a}.  The magnetic-optical trapping is extremely flexible
and able to simulate almost any geometry \cite{gre01,pet01}, as for example
carving out a deformed space in three or perhaps in smaller
dimensions.  One practical example is harmonic oscillator squeezing of
one coordinate from being fully allowed to completely
forbidden. Theoretically this is achieved by inserting an oscillator
potential with movable wall in one of the coordinates.  The practical
importance is especially due to the experimental availability.  To
reduce active dimensions by either simulating or directly squeezing
has been discussed before, f.ex. in Refs.
\cite{lev14,yam15,ros18,chr18,ros19,ros22,ros23}.

Practical investigations are obviously possible by adding an external
oscillator confinement to the otherwise ordinary treatment of relative
degrees-of-freedom.  This introduces necessarily one extra
three-dimensional degree-of-freedom on top of those used in the normal
description.  This does not sound, and is not, in any way prohibitive.
However, considering the complications added in going up one particle
number in a given system, as exemplified in the well-known two and
three-body systems, it is maybe desirable, or at least an advantage,
to minimize complications.

Recently, an equivalent method precisely offering this possibility has
been established.  The method was initially formulated \cite{nie01,gar19a},
developed and interpreted \cite{gar19b} in several publications,
where the case of $d$ varying between $2$ and $3$ especially is discussed for two and
three short-range interacting particles for both, bound \cite{gar19b,gar20a}
and continuum \cite{chr22,gar22} states. Particular emphasis has been 
made in the possibility of inducing the appearance of the Efimov
effect \cite{san18,gar21,gar21b,gar23}.

This $d$-method is
precisely as complicated as solving the corresponding relative two-
and three-body problems.  The formulation is in terms of non-integer
dimensions, which is a mathematical analytic continuation formulating
interpolation between integer dimensions.  The formulation provides a
translation between non-integer dimensions and external field constrained
three-dimensional calculations.  All necessary non-integer calculations
are without explicit use of the external confinement, but the results
can be interpreted as obtained with such a potential.  The translation
or interpretation extends to an approximate, deformed wave function
available for calculations of any desired observable, still obtained
entirely within this $d$-method.

In this report, we first formulate the $d$-method for an arbitrary
number of particles.  We derive general translations to the brute
force method of confining with an external deformed field. Obviously
the intrinsic many-body structure cannot be complicated, and ground
states are therefore most suited.  We specify to two and three
particles, where the assumptions can be substantially fewer, and compare
in details the specific transition between integer dimensions.

The paper is organized as follows. In Section 2 we describe the main
ingredients of the theoretical calculations with explicit inclusion of the
squeezing potential in three dimensions as well as for arbitrary values of 
the dimension. We also provide analytic solutions for both methods
assuming harmonic oscillator particle-particle interactions, and use
these results to derive an analytic relation between both methods
for different initial and final integer dimensions. In Section 3 we particularize
to the case of identical bosons. The results for two- and three-body
systems are shown in Section 4. We finish in Section 5 with the summary and
the conclusions.

\section{Formulation for $N$ particles}\label{sec2}

In this section we first give a formalism for an interacting system of 
$N$ particles, which are exposed to external one-body deformed oscillator 
potentials. The choice of parameters of the varying deformation decides the
transition between spatial dimensions. The goal is then to derive an
equivalent simpler method, the $d$-method, based on the use of non-integer
dimensions, and compare it with the brute force formulation using an external 
potential.

Once both methods have been presented, we solve them analytically by means of 
two-body harmonic oscillator particle-particle interactions. This permits a direct
comparison of the solutions from the two formulations, and we arrive at an
explicit formal translation between the two methods. Finally, these simple
results for different initial and final spatial dimensions are given.

In order to simplify the analysis of the numerical solutions, we are considering
only two-body short-range particle-particle potentials. In principle, we could
as well introduce three-body potentials into the game, but provided they are
of short-range character, they should not produce significant changes in the
conclusions. In any case, a careful investigation should be done to confirm this point. 

\subsection{General theory}\label{subsec2}

Let us consider $N$ particles with masses $m_i$, coordinates
$\bm{r}_i=(x_i,y_i,z_i)$, two-body interaction, $V_{ij}(\bm{r}_{ij})$,
between pair $i$ and $j$ with $\bm{r}_{ij}=\bm{r}_{i}- \bm{r}_{j}$,
and an external harmonic oscillator one-body interaction,
$V_\mathrm{ho}(\bm{r}_{i})$, acting on each of the particles.  The
Schr\"{o}dinger equation for the system is given by:
\begin{eqnarray} \label{eq40a}
  \bigg[- \sum_{i=1}^{N} \frac{\hbar^2}{2m_i} \Delta_{\bm{r}_i} +
    \sum_{i=1}^N V_\mathrm{ho}(\bm{r}_i) + \sum_{i<j} V_{ij}(\bm{r}_{ij}) - E_{N} \bigg] \Psi_\mathrm{tot} = 0 \; ,
\end{eqnarray}
where $E_{N}$ and $\Psi_{\mathrm{tot}}$ are the corresponding energy and wave function, and
$\Delta_{\bm{r}}$ is the Laplace kinetic energy operator related to
the vector $\bm{r}$.

Introducing the total mass of the system, $M=\sum_{i=1}^N m_i$, 
and the center-of-mass coordinate,
$\bm{R}_\mathrm{cm}=(X_\mathrm{cm},Y_\mathrm{cm},Z_\mathrm{cm})
=\frac{1}{M}\sum_{i=1}^N m_i \bm{r}_i$,
the external oscillator potential can be separated into
relative and center-of-mass parts, $V_\mathrm{ho}=V_\mathrm{ho}^\mathrm{rel}+V_\mathrm{ho}^\mathrm{cm}$.
The explicit division for this, possibly deformed, oscillator potential is:
\begin{eqnarray}  \label{eq60}
  V_\mathrm{ho} &=&\sum_{i=1}^N V_\mathrm{ho}(\bm{r}_i)=\frac{1}{2}  \sum_{i=1}^N m_i (\omega_{x}^2 x^2_i
  + \omega_{y}^2 y^2_i + \omega_{z}^2 z^2_i)  \\  \label{eq60a}
 V_\mathrm{ho}^\mathrm{cm} &=&  \frac{1}{2} M \big(\omega_{x}^2  X_{cm}^2 +
  \omega_{y}^2  Y_{cm}^2 + \omega_{z}^2  Z_{cm}^2\big) \;,  \\  \nonumber
  V_\mathrm{ho}^\mathrm{rel} &=& \frac{1}{2}  \sum_{i=1}^N m_i \big(\omega_{x}^2 (x_i-X_{cm})^2
  + \omega_{y}^2 (y_i-Y_{cm})^2 +  \omega_{z}^2 (z_i-Z_{cm})^2 \big) \\
   &=& \frac{1}{2} m (\omega_{x}^2  \rho_x^2 + \omega_{y}^2  \rho_y^2
  + \omega_{z}^2  \rho_z^2)  \; , \label{eq70}
\end{eqnarray}
which is only correct assuming the same oscillator
frequencies, $(\omega_{x},\omega_{y},\omega_{z})$, for each particle.

In Eq.(\ref{eq70}) we expressed the external relative oscillator potential in
terms of the separated components of the hyperradius coordinate,
$\rho$, where its length is defined by
\begin{eqnarray} \label{eq80}
  m  \rho^2 = \sum_{i=1}^N m_i (\bm{r}_{i}-\bm{R}_{cm})^2 =
 \sum_{i<j}  \frac{m_i m_j}{M} (\bm{r}_{i}- \bm{r}_{j})^2  \; .
\end{eqnarray}
The squared hyperradius can in turn be divided into the
spatial components used in Eq.(\ref{eq70}), that is 
\begin{eqnarray}
\lefteqn{
 \rho^2 = \rho_x^2 + \rho_y^2 + \rho_z^2  =
 } \label{eq90} \\ && 
 \sum_{i=1}^N \frac{m_i}{m}(x_i-X_\mathrm{cm})^2 + \sum_{i=1}^N \frac{m_i}{m}(y_i-Y_\mathrm{cm})^2 + 
 \sum_{i=1}^N \frac{m_i}{m}(z_i-Z_\mathrm{cm})^2\; . \nonumber
\end{eqnarray}
We emphasize that the hyperradius is not a vector, but expressed as a
length of different vectors.  The normalization $m$ can be arbitrarily
chosen and disappears in all observable quantities.

Eq.(\ref{eq40a}) can now be separated into intrinsic and center-of-mass
coordinates, which amounts to the following equations
\begin{eqnarray} \label{eq40b}
 \bigg(-\frac{\hbar^2}{2M} \Delta_{\bm{R}_\mathrm{cm}}+V_\mathrm{ho}^\mathrm{cm}-E_\mathrm{cm}\bigg)\Psi_\mathrm{cm}=0\; ,
\end{eqnarray}
\begin{eqnarray} \label{eq40}
  \bigg[- \sum_{i=1}^{N} \frac{\hbar^2}{2m_i} \Delta_{\bm{r_i}} +
  \frac{\hbar^2}{2M} \Delta_{\bm{R}_\mathrm{cm}} +  V_\mathrm{ho}^\mathrm{rel} 
  + \sum_{i<j} V_{ij}(\bm{r}_{ij}) - E_\mathrm{rel} \bigg] \Psi_\mathrm{rel} = 0 \; ,
\end{eqnarray}
where the energies and wave functions are related by 
$\Psi_\mathrm{tot} = \Psi_\mathrm{rel} \times \Psi_\mathrm{cm}$
and $E_{N} =  E_\mathrm{rel} + E_\mathrm{cm}$. It is important to note that 
the kinetic energy operator combination in Eq.(\ref{eq40}) can be
rewritten entirely in terms of relative coordinates.  The rigorous
proof would be rewriting in terms of Jacobi coordinates (any of the
possible sets of Jacobi coordinates is valid), and
simultaneously expressing Eq.(\ref{eq40}) in those coordinates. 
For details see Ref.~\cite{fab83}, where the precise definition of the $N$-body 
Jacobi coordinates can also be found.

The external squeezing oscillator potentials in Eq.(\ref{eq70}) are clearly
each zero for vanishing frequencies, whereas the other limit of
infinite frequency removes motion in the spatial directions.
Correspondingly, each infinite frequency leaves one less spatial
dimension. Practical examples are $\omega_{x} = \omega_{y}=0$ and $\omega_{z}$
varying from $0$ to $\infty$ corresponding to a transition from $3$ to
$2$ dimensions (3D to 2D), $\omega_{z} = 0$ with $\omega_{x} = \omega_{y}$ varying from $0$
to $\infty$ corresponding to a transition from 3D to 1D, and
$\omega_{z} = \infty$ with $\omega_{y} = 0$ and $\omega_{x}$ varying
from $0$ to $\infty$ corresponding to a transition from 2D to 1D.
The final dimension can even be zero, which would be reached when all
frequencies end up being infinitely large.

The squeezing variation from initial to final dimension proceeds by
increasing the energy in the squeezed direction from zero to infinity.
To reach the proper energy of the lower dimensional system, we have to
remove the corresponding zero-point energy, which is related
to the removed dimensions. This relative zero-point energy is
\begin{eqnarray} 
  E_{0} = \hbar \sum_{i=1}^{N-1} \Big(\omega_{x} (n_x(i) + 1/2)
 + \omega_{y} (n_y(i) + 1/2) +\omega_{z} (n_z(i) + 1/2) \Big) , \label{eq122} 
\end{eqnarray}
where the quantum numbers now refer to the relative
degrees-of-freedom.  The meaningful relative lower-dimensional energy,
$E_\mathrm{ext}=E_\mathrm{rel}-E_{0}$, therefore has to be the relative energy, $E_\mathrm{rel}$,
minus $E_{0}$, where the latter changes from zero to infinity during
the squeezing process.

For bosons the lowest quantum state is occupied by all $N$ particles
in the $N-1$ available relative degrees of freedom, that is
$n_x(i)=n_y(i)=n_z(i)=0$.  Then $E_0$ becomes
\begin{eqnarray} 
  E_0 =  \hbar  (\omega_{x} + \omega_{y} + \omega_{z}) (N-1)/2 \;.
\label{eq10}  
\end{eqnarray}

\subsection{The $d$-method}\label{subsubsec2}

The basis of the method is to consider the dimension $d$ as a parameter that
varies continuously between some initial and final integer dimension.
The external field is omitted as we are looking for another formulation with a
different phase-space that replaces the brute force use of a deformed
field producing the squeezing.

In order to obtain the $N$-body wave function of the system for a given 
value of the dimension $d$ we use the hyperspherical coordinates, which
in the present context only requires specification of the length,
$\rho$, defined in Eq.(\ref{eq80}).  All other coordinates, the
hyperangles ($\Omega$), are dimensionless variables. In this way it is possible to separate 
the complicated hyperangular dependence of the kinetic energy operator into
the hyperangular operator, $D_\Omega^2$, and the corresponding Schr\"{o}dinger
equation then reads:
\begin{eqnarray} \label{eq150}
 \bigg(-\frac{\partial^2}{\partial \rho^2}
 + \frac{\ell_{d,N}(\ell_{d,N}+1)}{\rho^2} + \frac{D^2_{\Omega}}{\rho^2} 
 + \frac{2m}{\hbar^2} \big(\sum_{i<j} V_{ij}(\bm{r}_{ij}) - E_{d}\big) \bigg)
 (\rho^{\ell_{d,N}+1}\Psi_{d}) = 0,
\end{eqnarray}
where  $\Psi_d$ is the total $d$-dimensional wave function. The pair interaction terms,
explicitly written in the equation, are understood as expressed in hyperspherical coordinates,
and the centrifugal barrier is given in terms of the  generalized angular momentum
$\ell_{d,N} = [(N-1)d -3]/2$.

 A large amount of possible
structures exist for $N$ particles.  This is a tremendous problem and
not addressed in general here, where we make simplifying assumptions.
In particular we assume relative $s$-waves between the particles, in such
a way that the dependence of $\Psi_d$ on the angles describing the direction
of the relative coordinates disappears. The wave function $\Psi_d$ can then 
be expanded as $\rho^{\ell_{d,N}+1}\Psi_d=\sum_K F_d^{(K)}(\rho) {\cal Y}_K^{(d)}(\alpha)$,
where ${\cal Y}_K^{(d)}(\alpha)$ are the $s$-wave hyperspherical harmonics in $d$ dimensions,
which depend only on the hyperangle $\alpha$ and the hypermomentum $K$, and 
whose explicit form can be found in the appendix of Ref.~\cite{gar20a}. Introducing
this wave function expansion into Eq.(\ref{eq150}), one easily gets that the radial functions,
$F_d^{(K)}$, can be obtained as the solution of the coupled set of equations:
\begin{eqnarray}
\lefteqn{\hspace*{-1cm}
\left[ -\frac{\partial^2}{\partial \rho^2}+\frac{\ell_{d,N}(\ell_{d,N}+1)+K(K+d(N-1)-2)}{\rho^2} - \frac{2mE_d}{\hbar^2}
\right]F_d^{(K)}(\rho)+
}
\nonumber \\ && \hspace*{3cm}
+\sum_{K'} \langle {\cal Y}_K(\alpha) | \sum_{i<j} V_{ij}(\bm{r}_{ij})| {\cal Y}_{K'}(\alpha)\rangle_\Omega F_d^{(K')}(\rho)=0,
\label{radeq}
\end{eqnarray}
where $\langle \rangle_\Omega$ indicates integration over the hyperangles, and where we have used that
$D_\Omega^2 {\cal Y}_K=K(K+d(N-1)-2){\cal Y}_K$ \cite{fab83}.

Although these expressions are in principle derived for integer values of the dimension, $d$,
which simply is the number of spatial degrees-of-freedom in our applications, we however
shall analytically continue all occurring functions to any desired non-integer value.

\subsection{Harmonic oscillator solutions}

Proceeding to obtain results from short-range potentials almost
necessarily requires numerical treatment no matter which method is
chosen.  This is true for even the simplest Gaussian two-body
interactions.  We postpone this application and pursue instead the
analytical insight obtained from harmonic oscillator potentials
between the particles.  The
obvious problem is here that oscillators are far from short-range
potentials, although a number of applications suggest the effectivity.
The reason is that the short-distance behavior is very similar for
short-range potentials and properly adjusted oscillators.

First we define the two-body interactions, $V_{ij}$, where the sum is the
combination entering in the Schr\"{o}dinger equation, Eq.(\ref{eq40}),
that is
\begin{eqnarray} 
  \sum_{i<j}V_{ij}(\bm{r}_{ij}) &=& \frac{1}{2} \omega_\mathrm{pp}^2
  \sum_{i<j} \frac{m_i m_j}{M} (\bm{r}_{i} - \bm{r}_{j})^2 =
 \frac{1}{2} m \omega_\mathrm{pp}^2 \rho^2 =   \;  \label{eq200} \\
 &=& \frac{1}{2} \omega_\mathrm{pp}^2  \sum_{i=1}^N m_i (\bm{r}_{i} - \bm{R}_{cm})^2 \;
 = \frac{1}{2}  \omega_\mathrm{pp}^2 \left(- M \bm{R}_{cm}^2 + \sum_{i=1}^N m_i \bm{r}_{i}^2\right) \nonumber
 \;, 
\end{eqnarray}
where Eq.(\ref{eq80}) has been used, and where $\omega_\mathrm{pp}$ describes the
overall strength of the given particle-particle interaction.  The mass
dependence of the individual two-body potential is not exactly the
reduced mass, since the sum of the two masses in the denominator is
missing.  The chosen form in Eq.(\ref{eq200}) allows expressing the
combined two-body interaction in terms of the hyperradius, and it also
allows the final expression in Eq.(\ref{eq200}) as a one-body
interaction, separating center-of-mass and relative coordinates.
For this it is essential to use the same frequency, $\omega_\mathrm{pp}$, for
all particle-particle interactions.  As we shall see, these subtle
choices in Eq.(\ref{eq200}) are essential to obtain the simple
analytical solutions, which on the other hand maintain the crucial
properties from short-range solutions.

We can now solve the external field relative equation in
Eq.(\ref{eq40}) with the oscillator interactions from
Eq.(\ref{eq200}).  The implicitly used Jacobi coordinates then
separates individually and in addition the three spatial coordinates
separate as well.  The center-of-mass motion is now completely
decoupled, and the effective oscillator frequencies obtained from 
the two-body interactions, Eq.(\ref{eq200}), and the external field, Eq.(\ref{eq70}), in the
three directions are $\sqrt{\omega_\mathrm{pp}^2 + \omega_{x}^2}$,
$\sqrt{\omega_\mathrm{pp}^2 + \omega_{y}^2}$, and $\sqrt{\omega_\mathrm{pp}^2 + \omega_{z}^2}$,
respectively.

Making use of Eq.(\ref{eq10}), we have that the energy $E_\mathrm{ext}=E_\mathrm{rel}-E_{0}$, relative to $\hbar \omega_\mathrm{pp}$
is then given by
\begin{eqnarray} \nonumber
  \frac{E_\mathrm{ext}}{\hbar \omega_\mathrm{pp}} &=&
   \bigg(\sqrt{1+\frac{\omega_{x}^2}{\omega_\mathrm{pp}^2}}
  - \frac{\omega_{x}}{\omega_\mathrm{pp}} \bigg) \sum_{i=1}^{N-1}(n_{x}(i)+1/2) \\
   \label{eq220}  &+& \bigg(\sqrt{1+\frac{\omega_{y}^2}{\omega_\mathrm{pp}^2}}
  - \frac{\omega_{y}}{\omega_\mathrm{pp}} \bigg) \sum_{i=1}^{N-1}(n_{y}(i)+1/2) \\
 \nonumber &+& \bigg(\sqrt{1+\frac{\omega_{z}^2}{\omega_\mathrm{pp}^2}}
  - \frac{\omega_{z}}{\omega_\mathrm{pp}} \bigg) \sum_{i=1}^{N-1}(n_{z}(i)+1/2)  \; ,
\end{eqnarray}
where the oscillator quantum numbers are related to the motion
relative to the center-of-mass divided into the three different
coordinate directions.

The corresponding $N$-body relative wave function, solution of Eq.(\ref{eq40}) with
the two-body interactions in Eq.(\ref{eq200}), is again the
oscillator solutions for the individual particles, decoupled in the
$x$, $y$ and $z$-directions, that is
\begin{equation}
\Psi_\mathrm{rel} = \prod_{i=1}^{N}  N_x H_{n_x(i)}(\frac{x_i}{b_{x,i}}) e^{-\frac{x_i^2}{2b_{x,i}^2}}
                              N_y H_{n_y(i)}(\frac{y_i}{b_{y,i}}) e^{-\frac{y_i^2}{2b_{y,i}^2}}
                              N_z H_{n_z(i)}(\frac{z_i}{b_{z,i}}) e^{-\frac{z_i^2}{2b_{z,i}^2}}
                              \label{eq320}
\end{equation}
where $N_x$, $N_y$, and $N_z$ are the normalization factors and the coordinates are the Jacobi coordinates in the
Schr\"{o}dinger equation.  The center-of-mass motion is not included, but it could 
be derived by solving Eq.(\ref{eq40b}). The
oscillator quantum numbers are related to the relative motion, and the
lengths in the different cases are defined as usual in terms
of the harmonic oscillator frequencies, that is
\begin{equation}
  b_{x,i}^2 = \frac{\hbar}{m_i \sqrt{\omega_{x}^2 +\omega_\mathrm{pp}^2}} \; ,
  b_{y,i}^2 = \frac{\hbar}{m_i \sqrt{\omega_{y}^2 +\omega_\mathrm{pp}^2}} \; ,
  b_{z,i}^2 = \frac{\hbar}{m_i \sqrt{\omega_{z}^2 +\omega_\mathrm{pp}^2}} \; .
  \label{eq330}
\end{equation}

Turning now to the $d$-method, the key equation of motion is now 
Eq.(\ref{eq150}).
The external field is substituted by the hyperspherical structure and
the $d$-dependence in the centrifugal barrier strength.  The angular
dependence would in general also contribute to the centrifugal
barrier, but our choice of interaction in Eq.(\ref{eq200}) leaves only
the harmonic oscillators second order dependence in hyperradius.  This
implies that all hyperspherical partial angular momenta are zero
except $s$-wave types, and only the spherical solution is left.
The energy is the energy of a three-dimensional oscillator with angular momentum,
$\ell_{d,N}$, and frequency, $\omega_\mathrm{pp}$, corresponding to 
\begin{eqnarray} 
 E_{d} = \hbar \omega_\mathrm{pp} (2n_r^{(d)} + \ell_{d,N} + \frac{3}{2})
 = \hbar \omega_\mathrm{pp} (2n_r^{(d)} + \frac{1}{2}(N-1) d) \; , \label{eq250}
 \end{eqnarray}
where the number of radial nodes, $n_r^{(d)}$, (multiplied by two) is
added to the generalized angular momentum quantum number introduced in 
Eq.(\ref{eq150}), and $\omega_\mathrm{pp}$ describes the oscillator pair
interaction.  The total radial wave function, $F_{d}(\rho)$, 
contained in $\Psi_d$, and solution of Eq.(\ref{radeq}), is
\begin{eqnarray} \label{eq255}
  F_{d}(\rho) = N_d e^{-\frac{\rho^2}{2 b_\mathrm{pp}^2}}
  L^{\ell_{d,N}+1/2}_{n_r^{(d)}}(\frac{\rho^2}{b_\mathrm{pp}^2}) \; ,
\end{eqnarray}
where $b_\mathrm{pp}^2= \hbar /(m\omega_\mathrm{pp})$ is the squared oscillator length,
$N_d$ is the normalization factor, that depends on $n_r^{(d)}$ and $\ell_{d,N}$,
and $L^{p}_{k}$ are the generalized Laguerre polynomials of order $k$.

\subsection{Translation between methods}

The above two methods are attempted constructed to be equivalent.
This means that computations in one of them can be translated and
interpreted in the parameters of the other method.  The calculations
are simplest with the $d$-method, whereas the external field method is
directly laboratory applicable.  Thus, it is the most efficient procedure to
calculate with $d$-parameters and translate the results to laboratory
variables.  

For equal energies, $E_\mathrm{ext} = E_{d}$, we have from Eqs.(\ref{eq220}) and (\ref{eq250}):
\begin{eqnarray}   \label{eq450}
 2n_r^{(d)} + \frac{1}{2}(N-1) d &=& 
  \bigg(\sqrt{1+\frac{\omega_{x}^2}{\omega_\mathrm{pp}^2}}
 - \frac{\omega_{x}}{\omega_\mathrm{pp}}\bigg) \sum_{i=1}^{N-1}(n_{x}(i)+1/2) \\ \nonumber
 &+& \bigg(\sqrt{1+\frac{\omega_{y}^2}{\omega_\mathrm{pp}^2}}
  - \frac{\omega_{y}}{\omega_\mathrm{pp}}\bigg) \sum_{i=1}^{N-1}(n_{y}(i)+1/2) \\ \nonumber
  &+&  \bigg(\sqrt{1+\frac{\omega_{z}^2}{\omega_\mathrm{pp}^2}}
  - \frac{\omega_{z}}{\omega_\mathrm{pp}}\bigg) \sum_{i=1}^{N-1}(n_{z}(i)+1/2) \;.
\end{eqnarray}
The dimension, $d$, can now be found from the combination of frequency
ratios and the quantum numbers of the state.  However, it is limited how
useful this relation really is, because the quantum numbers of the two
methods in general do not describe precisely the same state.  The
$d$-method is formulated as spherical and the kinetic energy accounted
for omitted hyperangles, that is all relative angular momentum
terms and all relative distance-related terms. Therefore, at best we
can compare average properties of angular momentum zero states.  This
is better indicated by rewriting Eq.(\ref{eq450}) as
\begin{eqnarray} \label{eq455}
  d = - \frac{4 n_r^{(d)}}{N-1}  
  &+& (2n_{x}^{(av)}+1) \bigg(\sqrt{1+\frac{\omega_{x}^2}{\omega_\mathrm{pp}^2}}
  - \frac{\omega_{x}}{\omega_\mathrm{pp}}\bigg) \\ \nonumber
     &+& (2n_{y}^{(av)}+1) \bigg(\sqrt{1+\frac{\omega_{y}^2}{\omega_\mathrm{pp}^2}}
  - \frac{\omega_{y}}{\omega_\mathrm{pp}}\bigg) \\ \nonumber
  &+& (2n_{z}^{(av)}+1) \bigg(\sqrt{1+\frac{\omega_{z}^2}{\omega_\mathrm{pp}^2}}
  - \frac{\omega_{z}}{\omega_\mathrm{pp}}\bigg) \; ,
\end{eqnarray}
where the average occupation numbers are given by $n_{x}^{(av)} = \frac{1}{N-1} \sum_{i=1}^{N-1}n_{x}(i)$,
$n_{y}^{(av)} = \frac{1}{N-1}\sum_{i=1}^{N-1}n_{y}(i)$, and $n_{z}^{(av)} = \frac{1}{N-1}\sum_{i=1}^{N-1}n_{z}(i)$,
respectively.

The frequencies $\omega_{x}$, $\omega_{y}$, and $\omega_{z}$, are
externally controlled, whereas $\omega_\mathrm{pp}$ is an artificial
oscillator frequency introduced to allow an analytical solution to be
compared with the numerical calculations.
However, whereas the realistic two-body interactions used in the calculations
are of short-range character, the harmonic oscillator potential is not. Therefore,
there is in principle no reason to expect a reasonable description of the system 
when the harmonic oscillator two-body potential is used. As we shall show in the
coming sections, this problem can in practice be solved by choosing $\omega_\mathrm{pp}$
such that it gives the same size for the system as in the calculations for
one of the dimensions, in particular for $d=2$, where an infinitesimal attraction 
always produces a bound state.

\section{The case of identical bosons}

Given an arbitrary $N$-body system, clearly very different structures may be chosen, 
and the $d$-method introduced above cannot describe them in the present simplified version.  The
connection in Eq.(\ref{eq455}) may at most approximately describe an average
translation for squeezing of a homogeneous state.  This may contain
some information, but we shall not pursue this possibility even though
average structures handled in this way could provide some insight.

Instead, we elaborate on the more comparable case, where all the particles
are in the same ground states. This therefore necessarily implies that we
are dealing with identical bosons, and all the quantum numbers are zero in 
the ground state. Also, all the masses, $m_i$, are identical, and 
the normalization mass, $m$, introduced in the definition of the hyperradius, Eq.(\ref{eq80}),
is also taken equal to the boson mass. Having all this in mind, we then have
\begin{eqnarray}   \label{eq465} 
 d =  \sqrt{1+\frac{\omega_{x}^2}{\omega_\mathrm{pp}^2}}-\frac{\omega_{x}}{\omega_\mathrm{pp}} 
  + \sqrt{1+\frac{\omega_{y}^2}{\omega_\mathrm{pp}^2}} - \frac{\omega_{y}}{\omega_\mathrm{pp}}
 +\sqrt{1+\frac{\omega_{z}^2}{\omega_\mathrm{pp}^2}} - \frac{\omega_{z}}{\omega_\mathrm{pp}} \;,
\end{eqnarray}
This expression is simple and, remarkably enough, independent of $N$.

The total ground state wave functions with the two methods are 
\begin{eqnarray}
\lefteqn{
\Psi_\mathrm{ext}=N_x N_y N_z \times
}  \label{eq520} \\ & &
   \exp\left[-\frac{1}{2} \sum_{i=1}^{N} \left( \frac{(x_i-X_{cm})^2}{b_x^2}
                              + \frac{(y_i-Y_{cm})^2}{b_y^2}
                              + \frac{(z_i-Z_{cm})^2}{b_z^2}\right)\right] \; , \nonumber
\end{eqnarray}
where the lengths in Eq.(\ref{eq330}) are now equal for all the particles.
From the $d$-method we get
\begin{eqnarray} 
\lefteqn{ 
  F_{d}(\rho) = N_d  e^{-\frac{\rho^2}{2 b_\mathrm{pp}^2}} = 
  } \label{eq555} \\ & & 
  N_d \exp\left[-\frac{1}{2} \sum_{i=1}^{N} \left(\frac{(x_i-X_{cm})^2}{b_\mathrm{pp}^2}
                              + \frac{(y_i-Y_{cm})^2}{b_\mathrm{pp}^2}
                              + \frac{(z_i-Z_{cm})^2}{b_\mathrm{pp}^2}\right)\right]
                              \nonumber
\end{eqnarray}
where Eq.(\ref{eq80}) has been used for identical particles and $m=m_i$. 

The wave
functions (\ref{eq520}) and (\ref{eq555}) can be made identical by proper scaling of the
coordinates according to their squeezing frequencies.  In particular, we can formulate this 
by interpreting the hyperradius, Eq.(\ref{eq90}), in the $d$-space as a hyperradius in the
ordinary three-dimensional space, but deformed along the $x$, $y$, and $z$ directions
by means of some scaling factor $s_x$, $s_y$, $s_z$.  In other words, the hyperradius is redefined
in the three dimensional space as
\begin{eqnarray} \label{eq567}
 <\rho^2> \longrightarrow <\rho_{x}^2> s_x^2 + <\rho_y^2> s_y^2 + <\rho_z^2> s_z^2 \;,
\end{eqnarray}  
and, after making equal Eqs.(\ref{eq520}) and (\ref{eq555}), we get that the scaling factors are
\begin{eqnarray} \label{eq565}
  s_x^2 = \frac{b_\mathrm{pp}^2}{b_x^2}\;,\;\;
  s_y^2 = \frac{b_\mathrm{pp}^2}{b_y^2}\;,\;\;
  s_z^2 = \frac{b_\mathrm{pp}^2}{b_z^2}\;,
\end{eqnarray}  
where $b_x^2$, $b_y^2$, and $b_z^2$ are given in Eq.(\ref{eq330}) with $m_i=m$, which leads to:
\begin{equation}
 \frac{1}{s_x^2}= \sqrt{1+\frac{\omega_{x}^2}{\omega_\mathrm{pp}^2}} \; , \;\;
 \frac{1}{s_y^2}= \sqrt{1+\frac{\omega_{y}^2}{\omega_\mathrm{pp}^2}} \; , \;\;
 \frac{1}{s_z^2}= \sqrt{1+\frac{\omega_{z}^2}{\omega_\mathrm{pp}^2}} \; . 
  \label{scale}
\end{equation}

Therefore, if the system is not squeezed along some direction, for instance direction $x$,
we then have that $\omega_x=0$ and therefore $s_x=1$, which implies that the space is not deformed
along that direction. On the contrary, if the system is fully confined
along one direction, say $\omega_x=\infty$, we get that $s_x=0$. In other
words, the scale parameters range between zero and one, $0\leq s_{x,y,z}\le 1$, 
where $s_{x,y,z}=0$ and $s_{x,y,z}=1$ mean, respectively, 
full confinement or no confinement along the given direction.
 
For these Gaussian wave functions related to the oscillator ground
state the translation is exact for the applied harmonic oscillator
interactions.  Using genuine short-range interactions vanishing at
large distances still allows this translation, where the pair
interaction frequency, $\omega_\mathrm{pp}$, must be appropriately adjusted.
Such an approximation is similar to the numerical approximation
obtained in a variational calculation using only one Gauss function.

The sizes of the oscillator solutions are controlled by $b_\mathrm{pp}^2 =\hbar/(m\omega_\mathrm{pp})$,
which has to be adjusted to reproduce
sizes obtained from realistic short-range two-body interactions.
It is perhaps remarkable that this interpretation and the derived
scalings are independent of both $N$ and $d$.

Summarizing, the spherical wave function obtained with the $d$-method depends on
$\rho$, which subsequently has to be deformed by scaling the different
directions. The ordinary three-dimensional wave function found in this
way is not restricted to the simple Gaussian form in
Eq.(\ref{eq520}).  The results from this translation provides the correct
long-range exponential fall-off for bound states. It is certainly much better than the Gaussian 
long-range behavior, but
how well it resembles the brute force calculated wave function has to
be separately evaluated.

\subsection{Varying initial and final dimension}

The connection between $d$ and the oscillator squeezing parameters is
given by the general expression in Eq.(\ref{eq465}), which involves
three independent squeezing frequencies corresponding to each dimension.  It may be
worth specifying each transition from initial to final dimension.
First we note the two limits
\begin{eqnarray}  \label{eq272}
 \sqrt{1+\frac{\omega_\mathrm{ho}^2}{\omega_\mathrm{pp}^2}} - \frac{\omega_\mathrm{ho}}{\omega_\mathrm{pp}}
 \rightarrow 1 \;\; {\rm for } \;\; \omega_\mathrm{ho} &=& 0 \;,\\  \label{eq274}
 \sqrt{1+\frac{\omega_\mathrm{ho}^2}{\omega_\mathrm{pp}^2}} - \frac{\omega_\mathrm{ho}}{\omega_\mathrm{pp}}
 \rightarrow 0 \;\; {\rm for } \;\; \omega_\mathrm{ho} &=& \infty \; ,
\end{eqnarray}
where $\omega_\mathrm{ho}=\infty$ or $0$ imply that the corresponding
dimensions are excluded or untouched throughout the process. If two or
three of these frequencies are equal the squeezing has cylindrical or
spherical symmetry. However, the squeezing from one dimension to
another does not have to be symmetric, and the parameter $d$ would
also vary in accordance with the frequency variation in
Eq.(\ref{eq465}).

When more than one direction is simultaneously squeezed, the
frequencies can all be either equal (symmetric) or unequal
(asymmetric).  For symmetric squeezing Eq.(\ref{eq465}) can be
rewritten as
\begin{eqnarray} \label{eq339}
 d = d_\mathrm{fin} + (d_\mathrm{ini}- d_\mathrm{fin})\bigg(\sqrt{1+\frac{\omega_\mathrm{ho}^2}{\omega_\mathrm{pp}^2}} 
 -  \frac{\omega_\mathrm{ho}}{\omega_\mathrm{pp}} \bigg) \;,
\end{eqnarray}
where $\omega_\mathrm{ho}=\hbar/(m b_\mathrm{ho}^2)$ is the squeezing frequency along all the squeezed directions, and 
$d_\mathrm{fin}$ and $d_\mathrm{ini}$ are, respectively, the final and initial integer dimensions.
Solving to express the squeezing frequency in terms of dimensions we get
\begin{eqnarray} \label{eq338}
  \frac{\omega_\mathrm{ho}}{\omega_\mathrm{pp}}  = \frac{b_\mathrm{pp}^2}{b_\mathrm{ho}^2} =
  \frac{(d_\mathrm{ini}-d)(d_\mathrm{ini}+d-2d_\mathrm{fin})}{2(d-d_\mathrm{fin})(d_\mathrm{ini}-d_\mathrm{fin})} \;,
\end{eqnarray}
which, when explicitly written for the possible symmetric transitions, becomes:
\begin{eqnarray}
    \frac{\omega_\mathrm{ho}}{\omega_\mathrm{pp}}  &=& \frac{b_\mathrm{pp}^2}{b_\mathrm{ho}^2} = \frac{(3-d)(d-1)}{2(d-2)}
    \;\;\;{\rm for}\;\;\;3\mbox{D} \rightarrow 2\mbox{D} \label{eq482} \\    
  \frac{\omega_\mathrm{ho}}{\omega_\mathrm{pp}}  &=& \frac{b_\mathrm{pp}^2}{b_\mathrm{ho}^2} = \frac{(3-d)(1+d)}{4(d-1)}
  \;\;\;{\rm for}\;\;\;3\mbox{D} \rightarrow 1\mbox{D} \label{eq483} \\ \label{eq484}
   \frac{\omega_\mathrm{ho}}{\omega_\mathrm{pp}} &=& \frac{b_\mathrm{pp}^2}{b_\mathrm{ho}^2} = \frac{(3-d)(3+d)}{6d}
  \;\;\;{\rm for}\;\;\;3\mbox{D} \rightarrow 0\mbox{D} \\  \label{eq485}
  \frac{\omega_\mathrm{ho}}{\omega_\mathrm{pp}} &=& \frac{b_\mathrm{pp}^2}{b_\mathrm{ho}^2} = \;\;\;\;\;\frac{(2-d)d}{2(d-1)}
  \;\;\;\;\;\;\;  {\rm for}\;\;\; 2\mbox{D} \rightarrow 1\mbox{D} \\ \label{eq486}
  \frac{\omega_\mathrm{ho}}{\omega_\mathrm{pp}} &=& \frac{b_\mathrm{pp}^2}{b_\mathrm{ho}^2} = \frac{(2-d)(d+2)}{4d}
  \;\;\;{\rm for}\;\;\; 2\mbox{D} \rightarrow 0\mbox{D} \\ \label{eq487}
  \frac{\omega_\mathrm{ho}}{\omega_\mathrm{pp}} &=& \frac{b_\mathrm{pp}^2}{b_\mathrm{ho}^2} = \frac{(1-d)(d+1)}{2d}
  \;\;\; {\rm for} \;\;\; 1\mbox{D} \rightarrow 0\mbox{D} \; .
\end{eqnarray}
We emphasize that these (inverse) relations only are possible for
symmetric squeezing, since $d$ otherwise depends on more than one
frequency ratio. Clearly this is the case when the transition covers
more than one dimension.

The symmetric squeezing implies that the involved scalings, $s_x$,
$s_y$ and $s_z$ correspondingly must be equal along the squeezed directions, 
whereas asymmetric
squeezing using Eq.(\ref{eq465}), also must lead to asymmetric
scaling.  The related wave functions in the transitions for these
different cases therefore necessarily must differ when reconstructed
in three dimensions.

\section{Numerical investigations}

For identical bosons in the ground state, the derived
relations between the non-integer dimension, $d$, and the
frequencies in both symmetric and asymmetric squeezing, Eq.(\ref{eq465}), are
independent of the particle number, $N$. The present spherical $d$-method
calculation is directly most applicable for ground states of a small
number of particles, where the dominant large-distance behavior is
universal.  The angular momentum barrier increases with $N$, and tends
to push the wave function to either smaller or larger
distances. Distances comparable to the constituent particle sizes is
problematic.  On the other hand, large distances are not allowed for
spherical bound states of large $N$, where the angular dependence of
the kinetic energy operator becomes important and effectively would
tend to split the system into clusters each with fewer particles.

In this report, we therefore concentrate on $N=2,3$, and analyze different
aspects of the transition between integer dimensions not investigated
in previous works.

\subsection{Two-body systems}

For two-body systems the 3D$\rightarrow$2D, 3D$\rightarrow$1D, and 2D$\rightarrow$1D 
transitions were investigated in detail in Refs.\cite{gar19a,gar19b}. In those works
three different Gaussian-shape and another three Morse-shape potentials were considered
to describe the interaction between two spinless particles. Following those works, we shall refer to them
as Potentials I, II, and III for both, Gaussian and Morse, shapes. The details about these six potentials 
can be found in Table~I of Ref.~\cite{gar19b}. It is interesting to note that their scattering
lengths in 3D range from modest values comparable to the range of the interaction 
(potentials I) to a large scattering length of about 30 times the potential range (potentials III).

In Ref.~\cite{gar19b} the connection between the dimension $d$ and the harmonic oscillator parameter of the
squeezing potential, $b_\mathrm{ho}$, was obtained by making equal the ground state energy of the system in both
calculations. The goal was then to establish universal relations between the squeezing harmonic oscillator
length and the dimension $d$ for each of the transitions investigated, as well as to provide a universal interpretation 
of the results. This was made basically through an analytic fit of the results, which in principle allows
predictions from non-integer dimension calculations of observables in trap experiments with external potentials.

In this work we have however derived analytic expressions connecting the squeezing harmonic
oscillator frequency, $\omega_\mathrm{ho}$ (or length $b_\mathrm{ho}$), and the dimension, $d$, by using a 
particle-particle harmonic oscillator potential with frequency $\omega_\mathrm{pp}$ (or length $b_\mathrm{pp}$).
The analytic relations for each possible transition are given in Eqs.(\ref{eq482}) to (\ref{eq487}).

The harmonic oscillator length of the particle-particle potential, $b_\mathrm{pp}$, can be easily related to
the root-mean-square (rms) radius of the system in 1D ($r_{\mbox{\tiny 1D}}$) or 2D ($r_{\mbox{\tiny 2D}}$),
since for a harmonic oscillator potential it is well known that $r_{\mbox{\tiny 1D}}=b_\mathrm{pp}/\sqrt{2}$ and  
$r_{\mbox{\tiny 2D}}=b_\mathrm{pp}$. In this way, by means of Eqs.(\ref{eq482}) to (\ref{eq487}), it is simple to express
the dimension $d$ as a function of $b_\mathrm{ho}/r_{\mbox{\tiny 1D}}$ or $b_\mathrm{ho}/r_{\mbox{\tiny 2D}}$.

In particular, here we focus on 3D$\rightarrow$2D, 3D$\rightarrow$1D, and 2D$\rightarrow$1D transitions, 
and by use of Eqs.(\ref{eq482}), (\ref{eq483}), and (\ref{eq485}) we get:
\begin{eqnarray}                                                                                                                                                                                                                                                         \frac{b_\mathrm{ho}}{r_{\mbox{\tiny 2D}}}&=&\sqrt{\frac{2(d-2)}{(3-d)(d-1)}} \mbox{ for 3D$\rightarrow$2D}, \label{rel32}\\
\frac{b_\mathrm{ho}}{r_{\mbox{\tiny 2D}}}&=&\sqrt{\frac{4(d-1)}{(3-d)(d+1)}} \mbox{ and }
\frac{b_\mathrm{ho}}{r_{\mbox{\tiny 1D}}}=\sqrt{\frac{8(d-1)}{(3-d)(d+1)}} \mbox{ for 3D$\rightarrow$1D}, \label{rel31} \\
\frac{b_\mathrm{ho}}{r_{\mbox{\tiny 2D}}}&=&\sqrt{\frac{2(d-1)}{d(2-d)}} \mbox{ and }
\frac{b_\mathrm{ho}}{r_{\mbox{\tiny 1D}}}=\sqrt{\frac{4(d-1)}{d(2-d)}} \mbox{ for 2D$\rightarrow$1D}. \label{rel21}
\end{eqnarray}

\begin{figure}[t]
\centering
\includegraphics[width=0.9\textwidth]{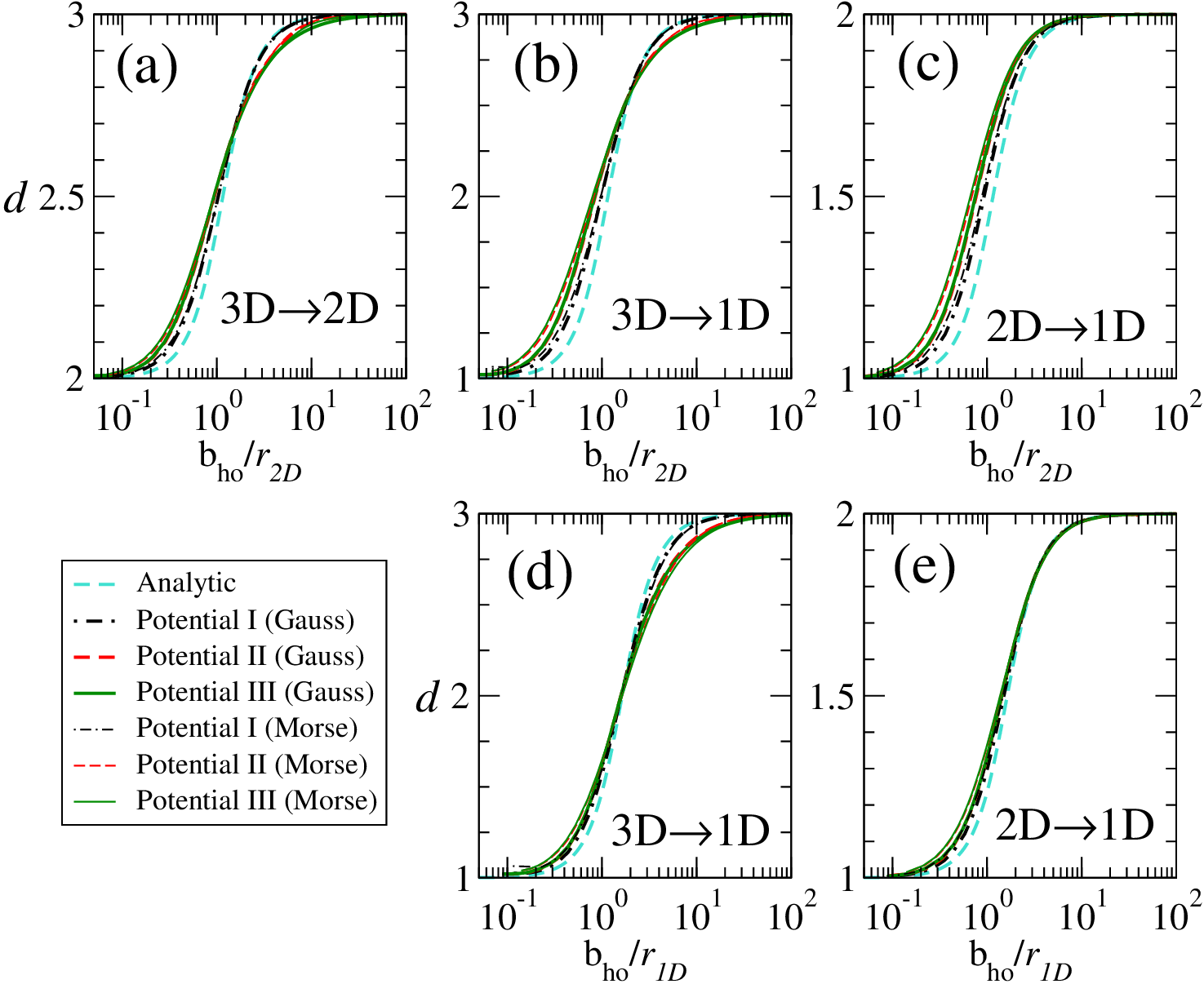}
\caption{Relation between the dimension $d$ and $b_\mathrm{ho}/r_{\mbox{\tiny 2D}}$ (upper row) or $b_\mathrm{ho}/r_{\mbox{\tiny 1D}}$
(lower row) for the different confinement transitions and potentials considered in this work. The dashed light-blue
curves correspond to the analytical estimates given in Eqs.(\ref{rel32}), (\ref{rel31}), and (\ref{rel21}). }
\label{fig1}
\end{figure}

In Fig.\ref{fig1} we compare the relation between $d$ and $b_\mathrm{ho}/r_{\mbox{\tiny 2D}}$, or $b_\mathrm{ho}/r_{\mbox{\tiny 1D}}$, 
obtained after the analytical estimates given in Eqs.(\ref{rel32}), (\ref{rel31}), and (\ref{rel21}) (dashed
light-blue curves) and the numerical calculations for the Gaussian (thick curves) and Morse potentials (thin curves)
mentioned above. The upper
row in the figure shows the relation between $d$ and $b_\mathrm{ho}/r_{\mbox{\tiny 2D}}$ for the 3D$\rightarrow$2D, 3D$\rightarrow$1D,
and 2D$\rightarrow$1D transitions (panels (a), (b), and (c)), and the lower row shows the relation between $d$ 
and $b_\mathrm{ho}/r_{\mbox{\tiny 1D}}$ for the 3D$\rightarrow$1D and 2D$\rightarrow$1D transitions (panels (d) and (e)).

Even though some differences between the numerical calculations can be seen, they appear mainly when comparing the results
obtained with both, Gaussian and Morse, potentials I and the rest of the calculations. This is a consequence of the fact
that potentials I have a relatively small scattering length, whereas the universal behavior of systems is related to the
presence of large scattering lengths. In any case, when normalizing the squeezing parameter $b_\mathrm{ho}$ with 
$r_{\mbox{\tiny 2D}}$ or  $r_{\mbox{\tiny 1D}}$, as we can see in the figure, the dependence on the value of the scattering length 
is very small in all the cases, especially in Fig.\ref{fig1}e. 

In Ref.\cite{gar19b} a clearly more universal connection between $d$ and $b_\mathrm{ho}$ was obtained (see Fig.11 in that reference). However, this was made after correcting $b_\mathrm{ho}$ with some scattering length dependent factor and using some very particular
length unit. Furthermore, the universal analytical curve was obtained by fitting the parameters in some very precise $d$-dependent function. All in all, although the method shown in Ref.\cite{gar19b} permits to relate $d$ and $b_\mathrm{ho}$, the procedure is not
very direct. For this reason, it is particularly interesting that the very simple method shown here provides a pretty much
universal connection between $d$ and $b_\mathrm{ho}$. Also, the numerical estimate given in Eqs.(\ref{rel32}), (\ref{rel31}), and (\ref{rel21}) follow reasonably well the numerical curves, without need of any numerical fit of the calculations.

At this stage it is interesting as well to investigate how the estimate given in Eq.(\ref{scale}) works for the
connection between the scale parameter deforming the three-dimensional wave function and the external squeezing potential.
Numerically this is done by choosing $s$ such that the overlap between the deformed three-dimensional $d$-wave function and the one obtained with
the external potential is maximum (and very close to 1, see Ref.\cite{gar19b}). 

Of course, in Eq.(\ref{scale}), $\omega_x$, $\omega_y$, and $\omega_z$ are non-zero only along the squeezed directions.
If we assume a symmetric squeezing for the 3D$\rightarrow$1D case ($\omega_x=\omega_y=\omega_\mathrm{ho}$), it is then not 
difficult to see that
\begin{equation}
\frac{1}{s^2}=\sqrt{1+\left(\frac{r_{\mbox{\tiny 2D}}}{b_\mathrm{ho}}\right)^4}
=\sqrt{1+4 \left(\frac{r_{\mbox{\tiny 1D}}}{b_\mathrm{ho}}\right)^4}
\label{sbrel}
\end{equation}
for all the three 3D$\rightarrow$2D, 3D$\rightarrow$1D, and 2D$\rightarrow$1D confinement processes.

\begin{figure}[t]
\centering
\includegraphics[width=0.9\textwidth]{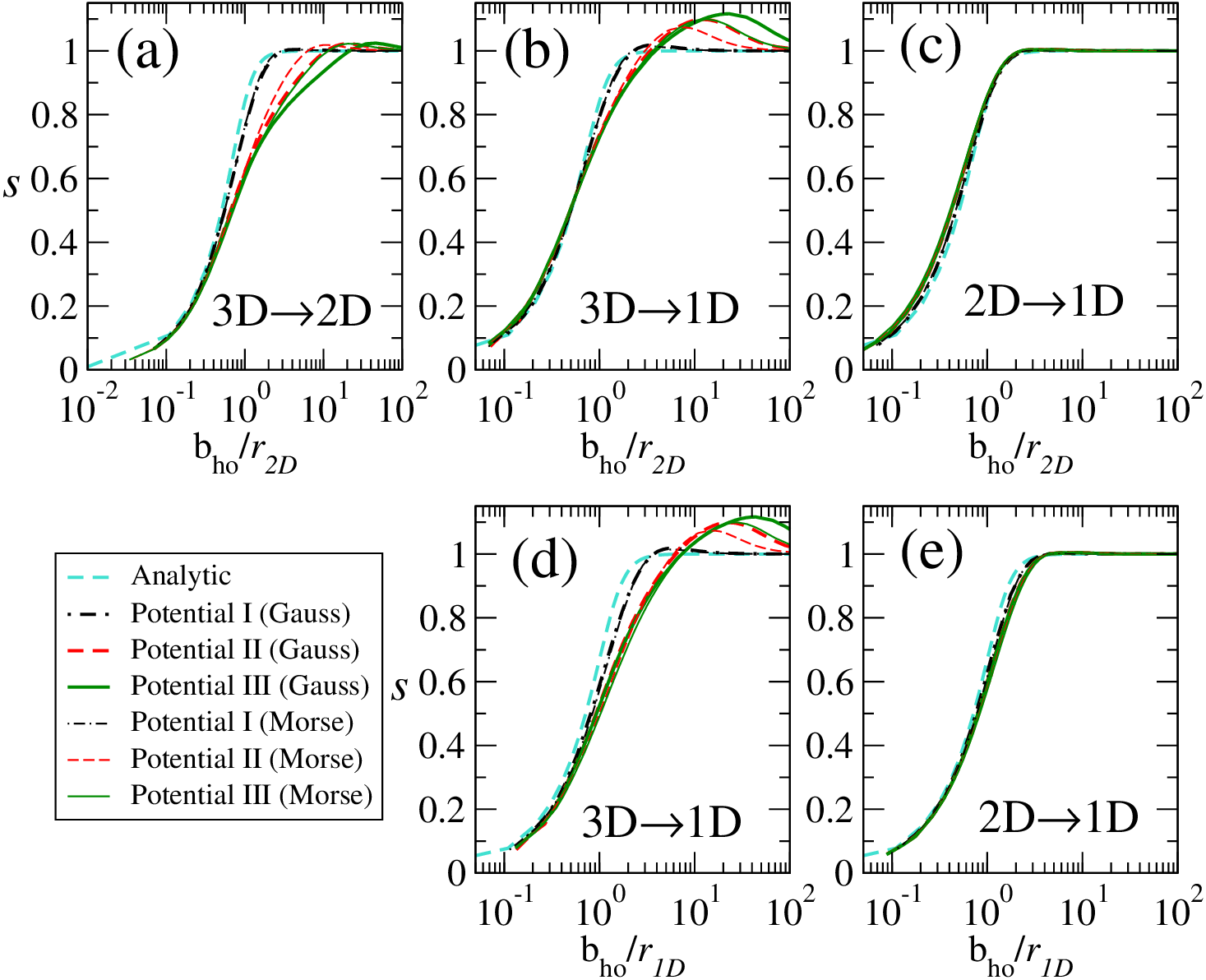}
\caption{Relation between the scale parameter $s$ and $b_\mathrm{ho}/r_{\mbox{\tiny 2D}}$ (upper row) or $b_\mathrm{ho}/r_{\mbox{\tiny 1D}}$
(lower row) for the different confinement transitions and potentials considered in this work. The dashed light-blue
curves correspond to the analytical estimates given in Eq.(\ref{sbrel}).}
\label{fig2}
\end{figure}

In Fig.~\ref{fig2} we show the connection between $s$ and $b_\mathrm{ho}/r_{\mbox{\tiny 2D}}$ (or $b_\mathrm{ho}/r_{\mbox{\tiny 1D}}$)
for all the cases considered in this work. The meaning of the curves is as in Fig.~\ref{fig1}. As discussed in Ref.~\cite{gar19b},
the computed scale parameter for small squeezing is sometimes bigger than 1. This mainly happens for the 
3D$\rightarrow$1D process. This is very likely the consequence of using a constant scale parameter.

As we can see in the figure the analytic expressions given in Eq.(\ref{sbrel}) work extremely well for the 
2D$\rightarrow$1D confinement process, as observed as well in Fig.~\ref{fig1}e. This is the case where the use of 
a constant scale parameter seems to be particularly appropriate. For the other cases the analytical results
work clearly better for small values of $b_\mathrm{ho}$ (large confinement), which are the most demanding cases
from the numerical point of view.

\subsection{Three-body systems}

In Ref.~\cite{gar20a} the case of three identical bosons was investigated for 3D$\rightarrow$2D confinement.
Two Gaussian and two Morse boson-boson potentials with small (potentials $A_g$ and $A_m$)
and large (potentials $B_g$ and $B_m$) scattering lengths in three dimensions were considered. The details of these potentials
can be found in Table~I of Ref.~\cite{gar20a}.

In this work we have generalized the procedure already described in Ref.~\cite{gar20a} for 3D$\rightarrow$2D confinement 
to transitions between any integer dimension. In fact, Eq.(\ref{eq482}) is Eq.(23) in Ref.~\cite{gar20a}.
For this reason here we shall concentrate on investigating to what extent the results shown in Ref.~\cite{gar20a}
for 3D$\rightarrow$2D are valid as well for other confinement processes. In particular we are going to 
focus on the (symmetric) 3D$\rightarrow$1D case using the same three-boson system as the one employed in Ref.~\cite{gar20a}.
We shall therefore maintain the same notation for the four boson-boson potentials used, potentials $A_g$ and $B_g$
(small and large scattering length with Gaussian shape) and potentials $A_m$ and $B_m$ (small and 
large scattering length with Morse shape).

\begin{figure}[t]
\centering
\includegraphics[height=0.34\textwidth]{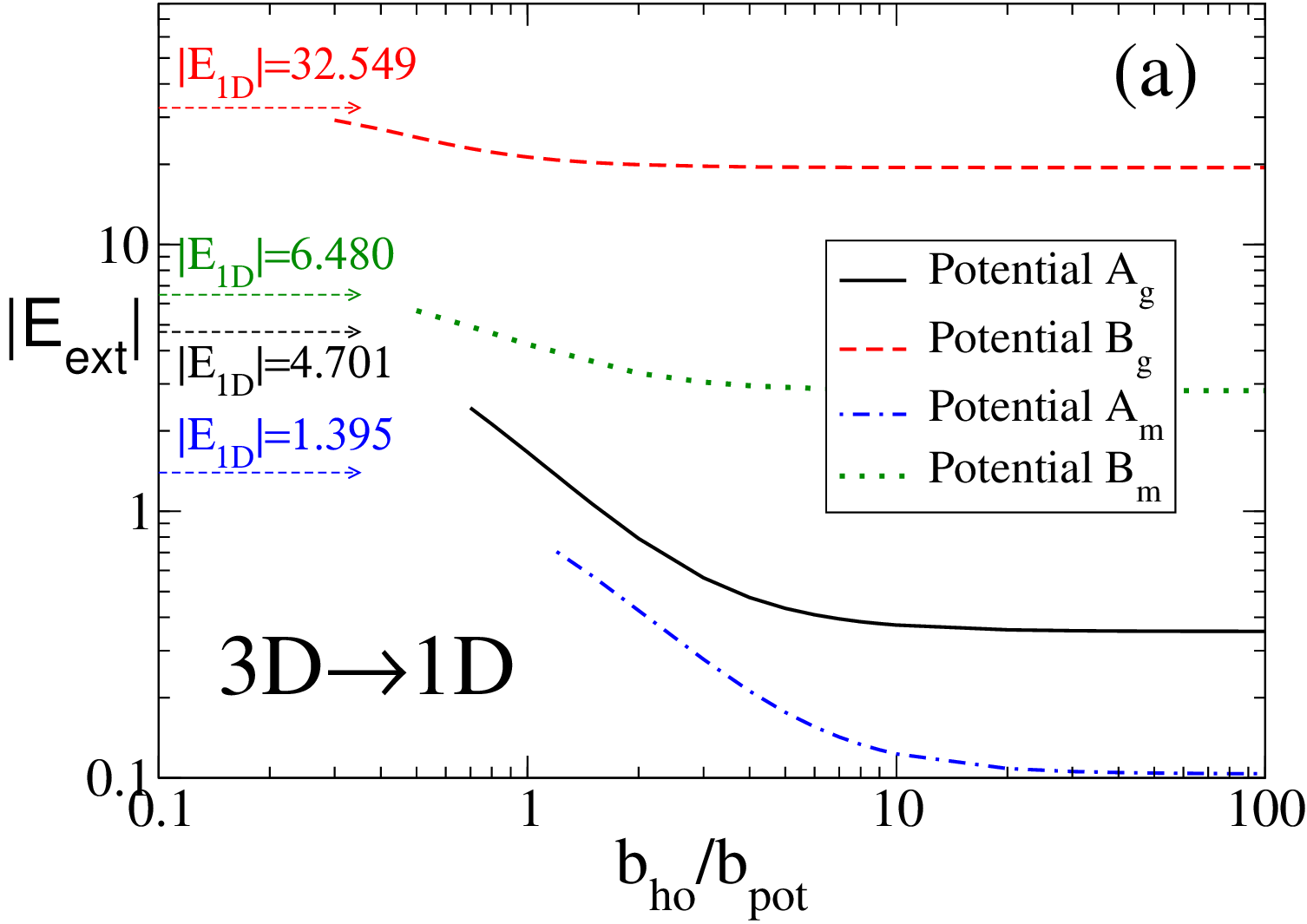}
\includegraphics[height=0.34\textwidth]{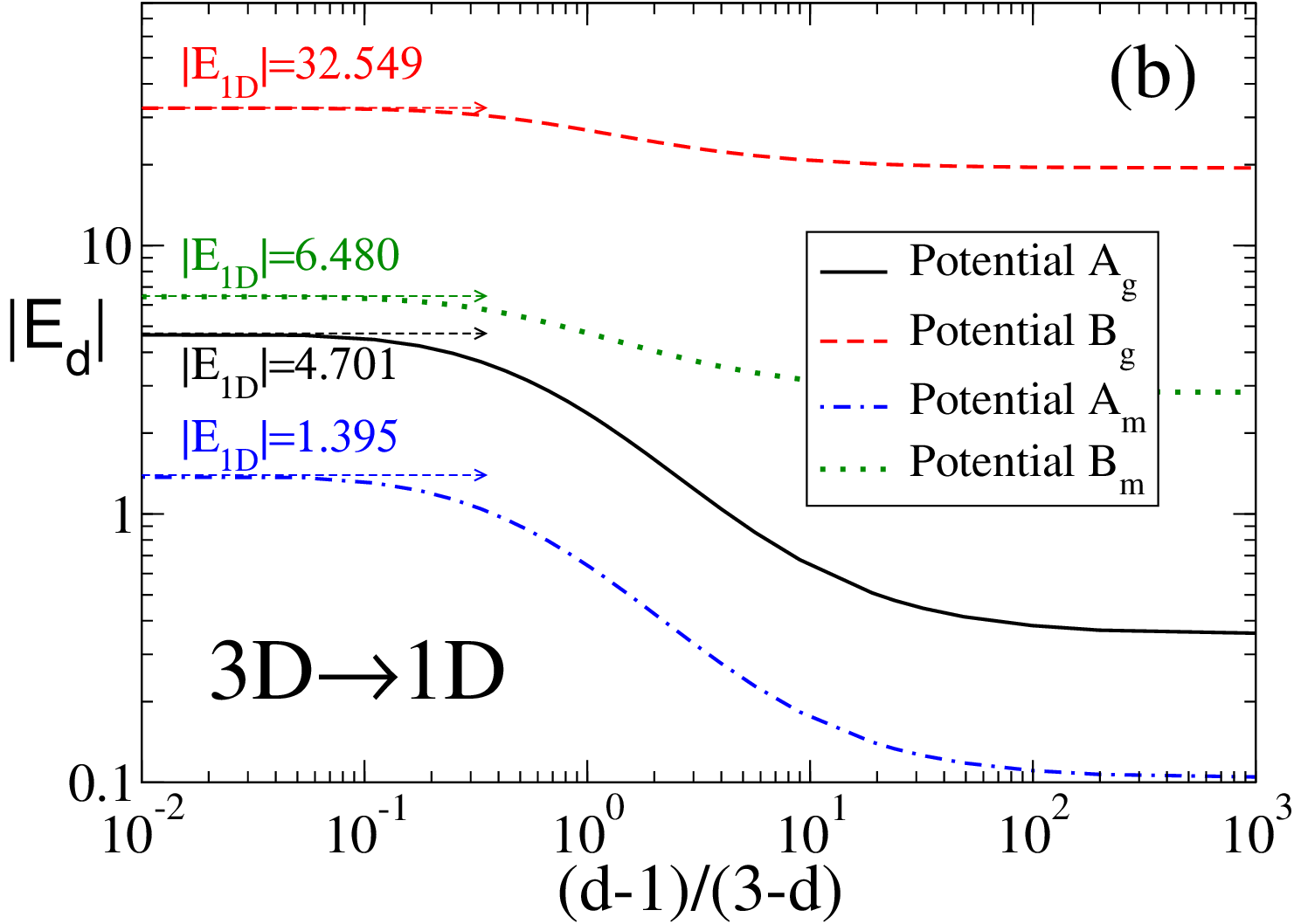}
\caption{For each of the Gaussian and Morse potentials used in this work, computed ground-state energies for 
3D$\rightarrow$1D confinement with an external squeezing
harmonic oscillator potential, panel (a), as a function of the harmonic oscillator length, $b_\mathrm{ho}$ (in units 
of the range of the interaction, $b_\mathrm{pot}$), and after
a calculation with the $d$-method, panel (b), as a function of $(d-1)/(3-d)$. The arrows indicate the computed
ground state energy in 1D for each of the potentials.}
\label{fig3}
\end{figure}

As already observed for 3D$\rightarrow$2D confinement, the three-body calculations with an external harmonic oscillator 
squeezing potential become more and more cumbersome when the system is progressively confined. In fact, for small values
of $b_\mathrm{ho}$ (large confinement) the three-body calculations become practically not doable. This is shown in Fig.~\ref{fig3}a, where we show,
for 3D$\rightarrow$1D confinement and for the four potentials used, the ground state energy of the system as a 
function of the squeezing harmonic oscillator length (in units of the range of the interaction, $b_\mathrm{pot}$). 
The arrows in the left part of the figure indicate the 
ground state energies obtained for each potential after a pure 1D calculation. As we can see, when $b_\mathrm{ho}$ decreases, the
computed energies approach the one corresponding to 1D, but it is not easy to go to smaller values of $b_\mathrm{ho}$ than
the ones shown in the figure. 

However, when the $d$-method is used, it is simple to perform the calculations using $d$ as a continuum parameter. There is
therefore no problem in reaching the 1D limit in this case. This is shown in Fig.~\ref{fig3}b, where the coordinate
in the abscissa axis is chosen as $(d-1)/(3-d)$, such that, as in Fig.~\ref{fig3}a, the right and left parts of the
figure correspond, respectively, to small and large confinement.
In this way, it is possible to relate $b_\mathrm{ho}$ and $d$ as those values
providing the same three-body ground state energy.  At least this is
true for the $b_\mathrm{ho}$ values for which the calculation with the
external potential are feasible.

\begin{figure}[t]
\centering
\includegraphics[height=0.37\textwidth]{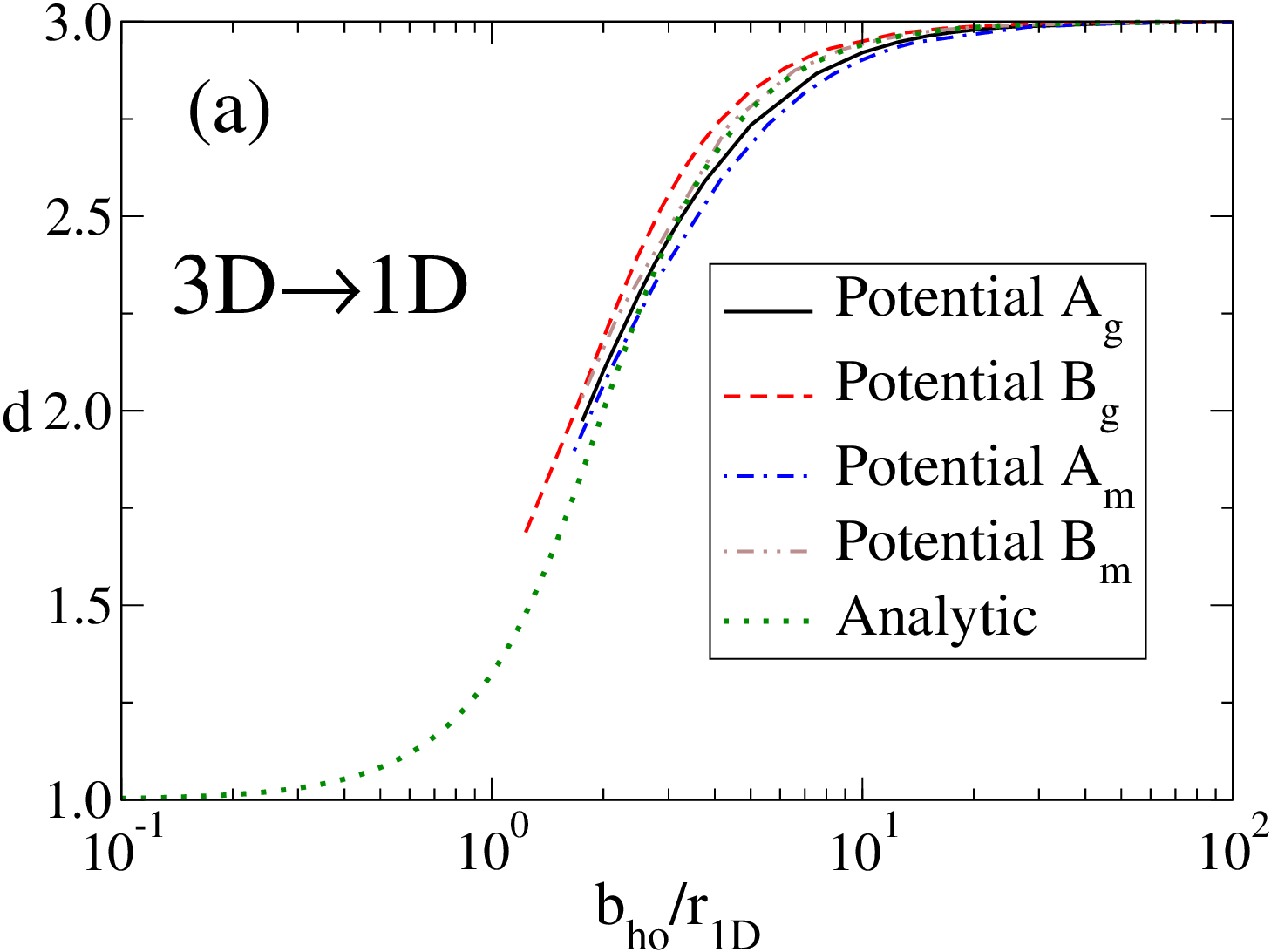}
\includegraphics[height=0.37\textwidth]{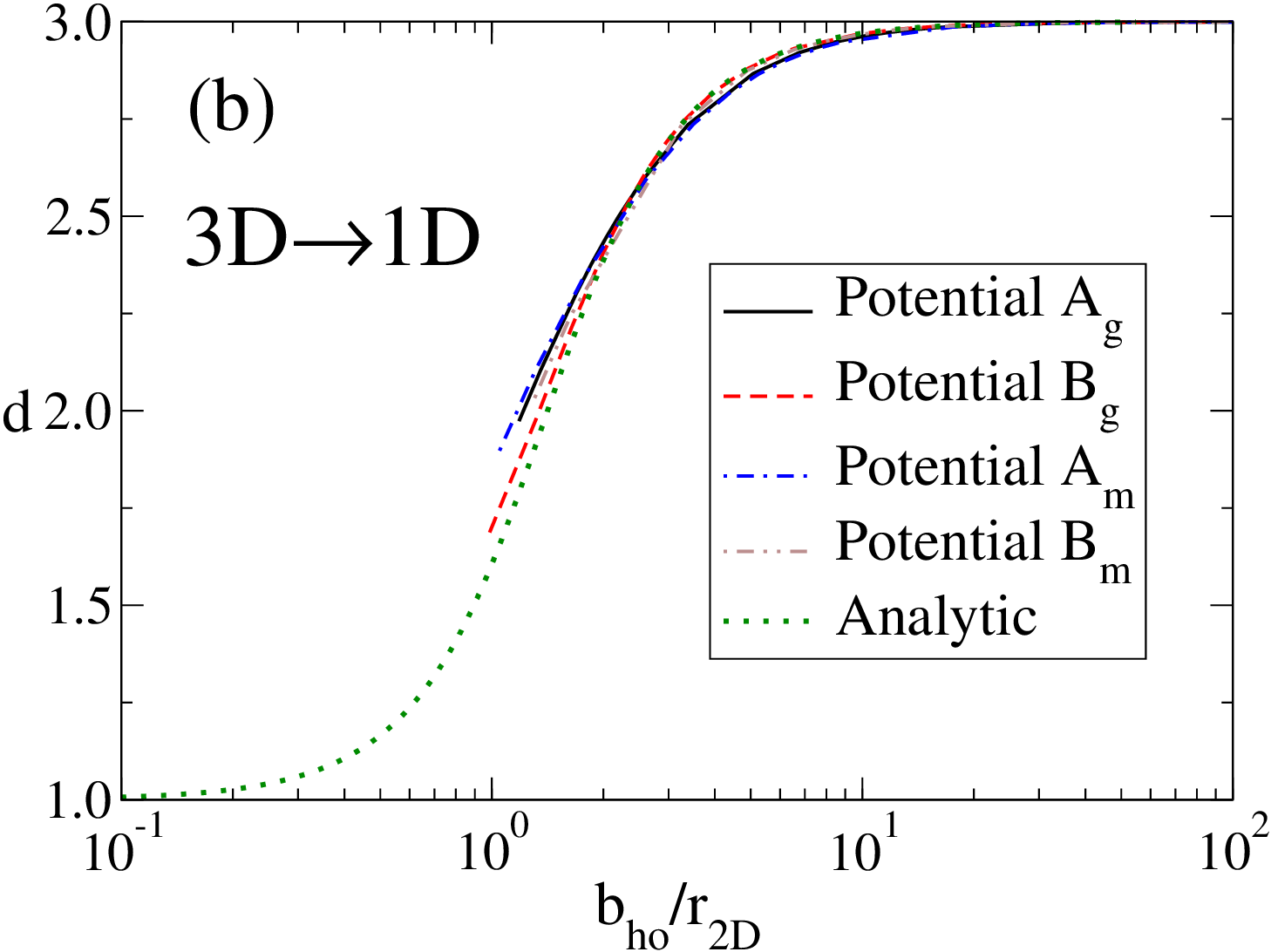}
\caption{For each of the Gaussian and Morse potentials used in this work, relation between the dimension $d$
and $b_\mathrm{ho}/r_{\mbox{\tiny 1D}}$, panel (a), and $b_\mathrm{ho}/r_{\mbox{\tiny 2D}}$, panel (b), obtained from the
values of $b_\mathrm{ho}$ and $d$ corresponding to the same ground state energy in Fig.\ref{fig3}. The dotted-green
curve are the analytic expressions given in Eq.(\ref{br3bd}).}
\label{fig4}
\end{figure}

When a harmonic oscillator with frequency $\omega_\mathrm{pp}$ (or length $b_\mathrm{pp}$) is used as boson-boson potential,
it is known that the rms radius of the three-body system in 2D and 1D are given, respectively, by
$r_{\mbox{\tiny 2D}}=b_\mathrm{pp}\sqrt{2/3}$ and $r_{\mbox{\tiny 1D}}=b_\mathrm{pp}\sqrt{1/3}$. We then have that Eq.(\ref{eq482})
can be rewritten as 
\begin{equation}
\frac{b_\mathrm{ho}}{r_{\mbox{\tiny 2D}}}= \sqrt{\frac{3(d-2)}{(3-d)(d-1)}},
\end{equation}
which is Eq.(38) of Ref.~\cite{gar20a}, valid for 3D$\rightarrow$2D processes, and, similarly, Eq.(\ref{eq483})
can be rewritten as
\begin{equation}
\frac{b_\mathrm{ho}}{r_{\mbox{\tiny 2D}}}= \sqrt{\frac{6(d-1)}{(3-d)(d+1)}} \mbox{ and }
\frac{b_\mathrm{ho}}{r_{\mbox{\tiny 1D}}}= \sqrt{\frac{12(d-1)}{(3-d)(d+1)}},
\label{br3bd}
\end{equation}
which provides an analytic form for the relation between $d$ and $b_\mathrm{ho}$ valid for 3D$\rightarrow$1D confinement.

In Fig.~\ref{fig4} we show, for each of the Gaussian and Morse potentials used in this work, the relation between $d$
and $b_\mathrm{ho}$ obtained from the values that make equal the ground state energies in Fig.\ref{fig3}. In the panels (a)
and (b) we show $d$ versus $b_\mathrm{ho}/r_{\mbox{\tiny 1D}}$ and versus $b_\mathrm{ho}/r_{\mbox{\tiny 2D}}$, respectively. We can see 
that in both cases the computed curves tend to merge into a single curve, although this is clearly more evident
in Fig.~\ref{fig4}b. It seems therefore that $r_{\mbox{\tiny 2D}}$ is a more convenient length unit in order to obtain
a universal behavior. This same length unit is also giving rise to a pretty much universal curve in the
3D$\rightarrow$2D case, as shown in Fig.~5 of Ref.~\cite{gar20a}. 

In Fig.~\ref{fig4} we show as well the analytic relation between $d$ and $b_\mathrm{ho}/r_{\mbox{\tiny 2D}}$ and 
$b_\mathrm{ho}/r_{\mbox{\tiny 1D}}$ given in Eq.(\ref{br3bd}), obtained assuming a harmonic oscillator boson-boson 
interaction. The result is shown by the dotted-green curves. We can see that in both panels the analytic
curves follow very closely the ones obtained numerically, although, again, using $b_\mathrm{ho}/r_{\mbox{\tiny 2D}}$
as coordinate, Fig.~\ref{fig4}b, seems to be clearly better. We can therefore consider the first equality in 
Eq.(\ref{br3bd}) as a very good estimate of the relation between $d$ and $b_\mathrm{ho}$ pretty much 
reliable for large confinement scenarios, where the numerical calculations with the external field
are extremely complicated. 

\begin{figure}[t]
\centering
\includegraphics[height=0.36\textwidth]{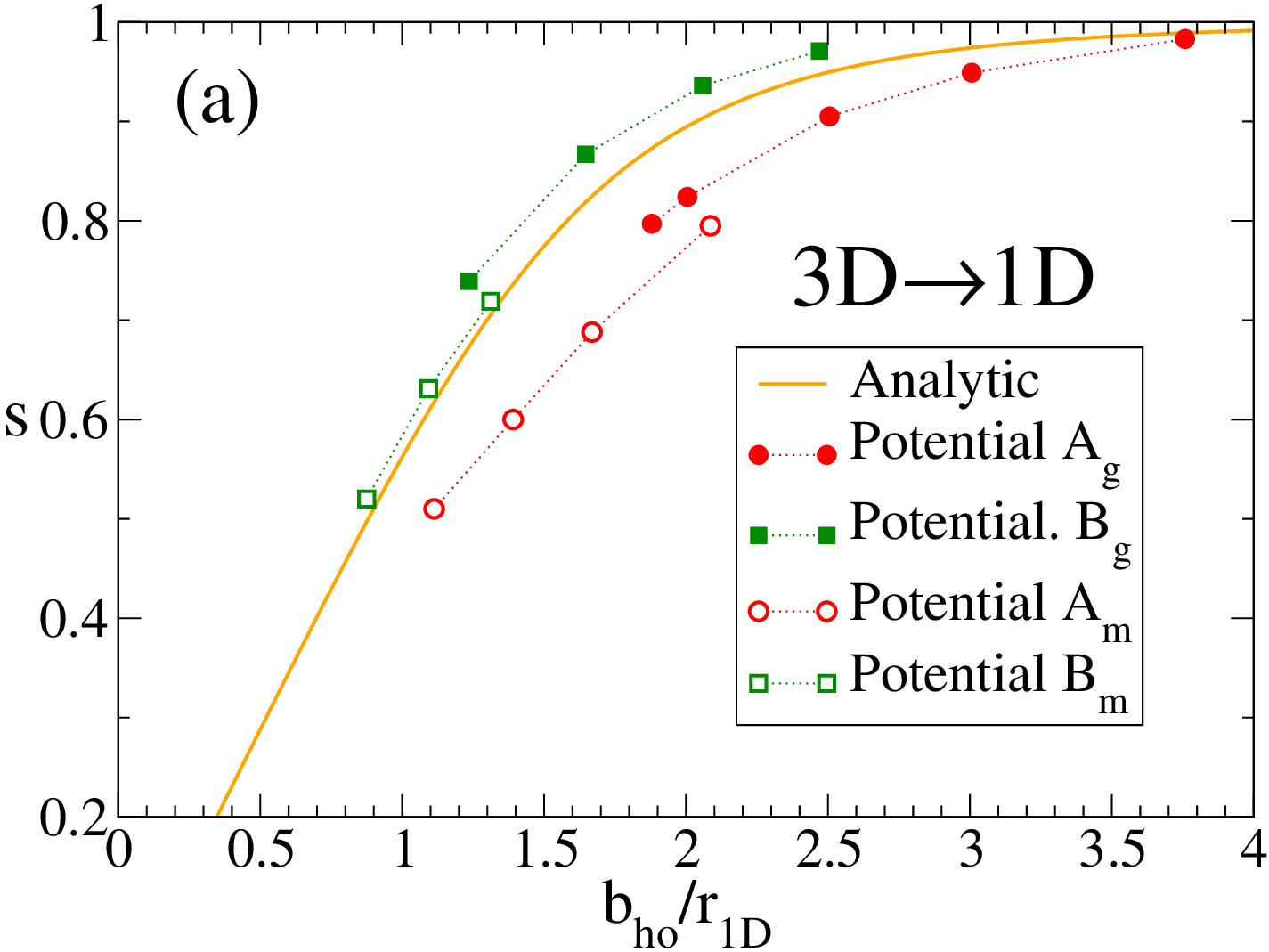}
\includegraphics[height=0.36\textwidth]{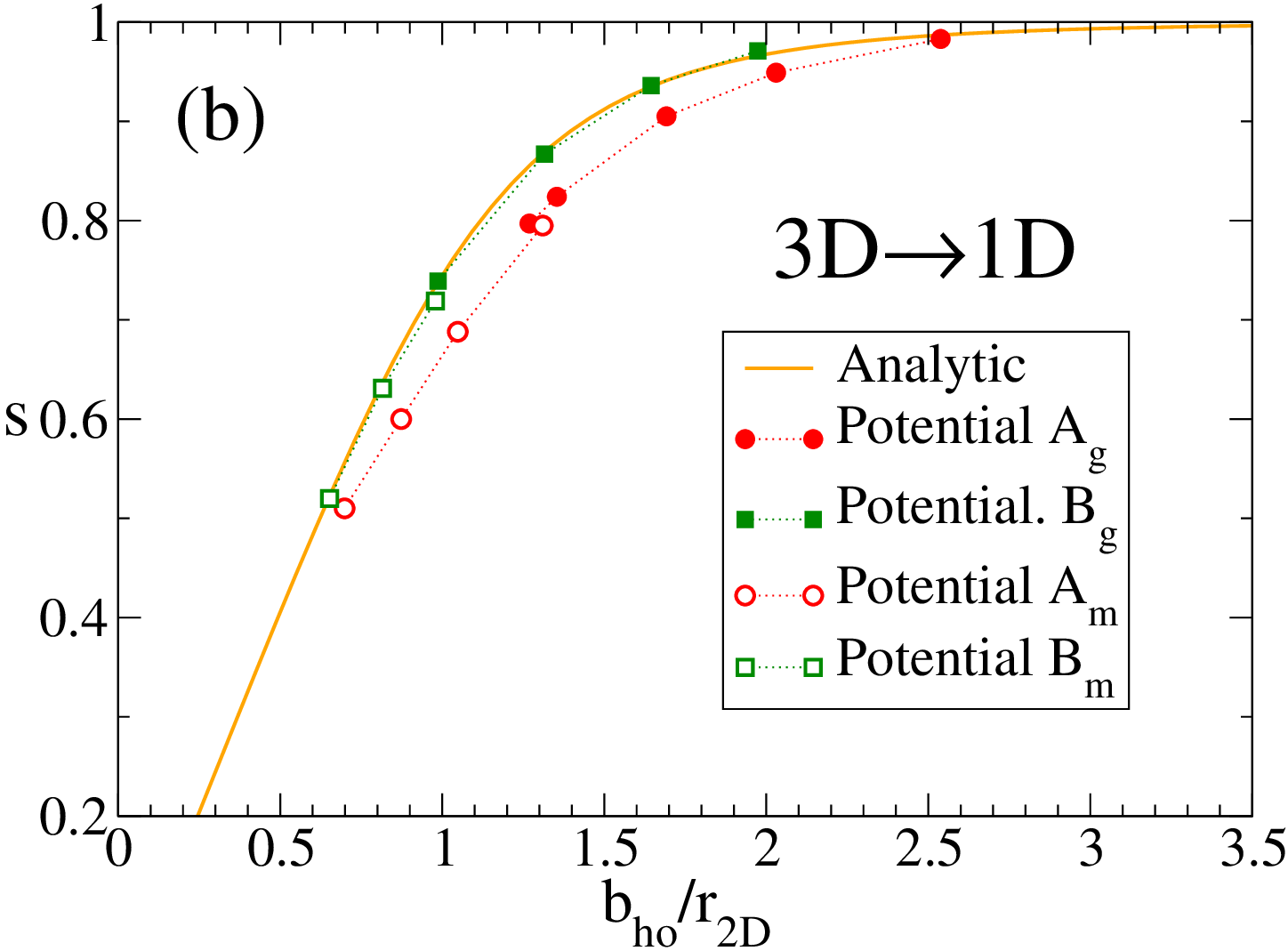}
\caption{For each of the Gaussian and Morse potentials used in this work, relation between the scale parameter $s$
and $b_\mathrm{ho}/r_{\mbox{\tiny 1D}}$, panel (a), and $b_\mathrm{ho}/r_{\mbox{\tiny 2D}}$, panel (b). The solid-orange
curve are the analytic expressions given in Eq.(\ref{sbrel2}).}
\label{fig5}
\end{figure}

Let us finally focus on the scale parameter $s$ introduced in Eq.(\ref{eq565}), which permits
to interpret the $d$-wave function as an ordinary three-dimensional wave function deformed along the
squeezing directions. As discussed in Ref.~\cite{gar20a}, the scale parameter can be obtained 
as the value of $s$ that maximizes the overlap between both wave functions, the one obtained after the calculation 
with the external field and the one obtained after the $d$-calculation (and interpreted as deformed
along the squeezing directions). Also, with Eq.(\ref{scale}) and symmetric confinement for the
3D$\rightarrow$1D case, we can see that the scale parameter, $s$, and
$b_\mathrm{ho}$ are related for oscillator interactions by the analytic
expressions
\begin{equation}
\frac{1}{s^2}=\sqrt{1+\frac{9}{4}\left(\frac{r_{\mbox{\tiny 2D}}}{b_\mathrm{ho}}\right)^4}
=\sqrt{1+9 \left(\frac{r_{\mbox{\tiny 1D}}}{b_\mathrm{ho}}\right)^4},
\label{sbrel2}
\end{equation}
where we have again used that $r_{\mbox{\tiny 2D}}=b_\mathrm{pp}\sqrt{2/3}$ and $r_{\mbox{\tiny 1D}}=b_\mathrm{pp}\sqrt{1/3}$, and
where the first equality is the one already given in Eq.(39) of Ref.~\cite{gar20a}.

In Fig.~\ref{fig5} the closed and open circles and the closed and open squares show, for the potentials $A_g$,
$A_m$, $B_g$, and $B_m$, respectively, the relation between the scale parameter $s$ and 
$b_\mathrm{ho}/r_{\mbox{\tiny 1D}}$, panel (a), and $b_\mathrm{ho}/r_{\mbox{\tiny 2D}}$, panel (b), obtained after the numerical 
calculation for 3D$\rightarrow$1D confinement.
In both panels the solid-orange curve is the analytic estimate given in Eq.(\ref{sbrel2}). We again see that the
$b_\mathrm{ho}/r_{\mbox{\tiny 2D}}$ coordinate, panel (b), is clearly more appropriate than 
$b_\mathrm{ho}/r_{\mbox{\tiny 1D}}$, panel (a). In fact, when the $r_{\mbox{\tiny 2D}}$ length unit is chosen
the $B$-potentials follow extremely well the analytic form given in Eq.(\ref{sbrel2}). For the $A$-potentials
the disagreement with the analytic expression is clearly visible. The result in Fig.~\ref{fig5}b is 
very similar to the one obtained in Ref.~\cite{gar20a} for the 3D$\rightarrow$2D case, and, as discussed in 
this reference, the disagreement between the analytic expression and the results for the $A$-potentials
can be a consequence of the use of constant scale parameter combined with the fact that the 
$A$-potentials are the ones corresponding to very large scattering length, which could make them
much more sensitive to the confinement.

Therefore, the first analytic relation in Eq.(\ref{sbrel2}) can be
safely used to extract the connection between the scale parameter,
$s$, and $b_\mathrm{ho}$ for small values of $b_\mathrm{ho}$, i.e. for large
confinements.  This is precisely the situation where the numerical
calculations with the external squeezing field is most complicated.
Furthermore, this conclusion is, independently of the potential, also
reached in Ref.~\cite{gar20a}, and similar to the agreement found for
the first analytic expression in Eq.(\ref{br3bd}).

\section{Summary and conclusions}

We formulate two methods to continuously change the spatial dimension
for a system of $N$ particles. The first method, brute force method,
employs an external varying one-body deformed oscillator
potential. Allowing the oscillator length to take values from zero to
infinity, we reduce one or more dimensions from fully present to
completely removed from the available space. The two-body
particle-particle interaction is of short range.  In principle we can
study the behavior as the spatial dimension is reduced.  The solution
involves both relative and center-of-mass degrees-of-freedom and
depends clearly on the number of particles and the interactions.

The second method, the $d$-method, has no external potential, only the
hyperradial equation is considered for $s$-waves, where a centrifugal
barrier depends on $N$ and the dimension parameter, $d$.  Only
relative motion is then appearing for the short-range interacting
particles.  The process of reducing the dimension is achieved by
varying the parameter $d$.

To gain insight, we assume the particle-particle interaction is a
possibly deformed harmonic oscillator potential with coordinate dependent
frequencies, which however must be the same for all particles.  These
oscillator solutions for both methods are now analytic, and found for arbitrary
potentials of this oscillator form and any value of $N$. Insisting on identical
energies for the two methods, we arrive at a simple analytic connection between
the dimension parameter, $d$, and the frequencies of the external
one-body potential. We give explicitly the analytic formula for all
possible dimension transitions. Furthermore, a similar relation can also be obtained between the
squeezing harmonic oscillator frequency and the scale parameters that describe
the deformation of the $d$-wave function when translated back into the ordinary
three-dimensional space. We scale the three
coordinates from the spherical $d$-method, and obtain a wave function
very similar to that of the brute force method.  The scaling sizes are
therefore also given as function of the confining oscillator frequencies.

The generalization to use of genuine short-range potentials is
achieved by insisting on identical energies for $d=2$ from this
realistic interaction and the two-body harmonic oscillator.  The
analytical relation between $d$ and the confinement potential is then
modified by substituting the realistic mean-square-radius for the same
quantity in the particle-particle oscillator calculation.
These results are valid for any $d$ and any $N$.

When the system under consideration is made of identical bosons they can all occupy the lowest quantum
state. The consequence is that the expressions derived present the nice feature of being independent 
of the number of particles.

In this work, the equivalence between the two methods and the validity of the analytic expressions 
derived have been tested for the cases of two and three identical bosons. The numerical calculations 
and two-body potentials employed are as shown in previous works for two and 
three-body systems. At the two-body level we have investigated the 
transitions with final dimension 1 or 2, i.e., 3D$\rightarrow$2D, 3D$\rightarrow$1D, and 2D$\rightarrow$1D.
For three identical bosons we have focused on the 3D$\rightarrow$1D process. For both, two- and three-body
systems, we have assumed a symmetric confinement in 3D$\rightarrow$1D case.

From the calculations and the comparison with the derived analytic relations we have observed in all
the cases a nice agreement in both, the two- and three-boson cases. We have found that the function
describing the relation between the oscillator length of the squeezing potential and the dimension $d$
shows a pretty much universal behaviour provided that the harmonic oscillator
length is normalized with the root-mean-square radius of the system in two dimensions, $r_{\mbox{\tiny 2D}}$. 
This is like this for two-body and three-body systems, and agrees with the results shown previously for three
identical bosons and 3D$\rightarrow$2D squeezing. For the relation between the squeezing oscillator length
and the scale parameter, $s$, the use of $r_{\mbox{\tiny 2D}}$ as unit length gives rise as well to a universal curve, 
especially in the large squeezing region. The overall agreement is remarkably good, especially for
not too small scattering lengths, where the short-range properties are more important.

In conclusion, we have provided a simple method, the $d$-method, that permits to perform simple numerical
calculations, no matter the $d$-value, equivalent to confinement processes with an external field, which are
much more complicated and close to impossible for large confinement scenarios.
We have also derived analytic expressions that relate both descriptions of the same process, in such a way
that it is most effective to perform the simple $d$-method calculations for
non-integer dimensions, and afterwards translate to find the observables in ordinary three dimensional space.

\bmhead{Acknowledgements}

This work has been partially supported by Grant PID2022-136992NB-I00 funded by MCIN/AEI/10.13039/501100011033 and by ``ERDF A way of making Europe''.


%\bibliography{bibfile}% common bib file

\begin{thebibliography}{10}

\bibitem{pet00} D.S. Petrov, M. Holzmann, and G.V. Shlyapnikov, Phys. Rev. Lett. 84, 2551 (2000).

\bibitem {gre01} M. Greiner, I. Bloch, O. Mandel, T. W. H\"{a}nsch, and T.Esslinger, Appl. Phys. B 73, 769 (2001).

\bibitem{blo05} I. Bloch, Nature Phys. 1, 23 (2005).

\bibitem{nis08} Y. Nishida and S. Tan, Phys. Rev. Lett. 101, 170401 (2008).

\bibitem{gar19a}
E. Garrido, A.S. Jensen, and R. \'{A}lvarez-Rodr\'{\i}guez, Phys. Lett. A 383, 2021 (2019).

\bibitem{pet01}
D.S. Petrov and G.V. Shlyapnikov, Phys. Rev. A 64, 012706 (2001).

\bibitem{lev14} J. Levinsen, P. Massignan, and M.M. Parish, Phys. Rev. X 4, 031020 (2014).

\bibitem{yam15} M.T. Yamashita, F.F. Bellotti, T. Frederico, D.V. Fedorov, A.S. Jensen, N. T. Zinner, J. Phys. B: At. Mol. Opt. Phys. 48,  025302 (2015).

\bibitem{ros18} D.S. Rosa, T. Frederico, G. Krein, and M.T. Yamashita, Phys. Rev. A 97, 050701(R) (2018).

\bibitem{chr18} E. R. Christensen,  A.S. Jensen, and E. Garrido, Few-Body Syst 59, 136 (2018).

\bibitem{ros19} D.S. Rosa, T. Frederico, G. Krein, and M.T. Yamashita, J. Phys. B: At. Mol. Opt. Phys. 52,  025101 (2019).

\bibitem{ros22} D.S. Rosa, T. Frederico, G. Krein, and M.T. Yamashita, Phys. Rev. A 106, 023311 (2022).

\bibitem{ros23} D.S. Rosa, T. Frederico, G. Krein, and M.T. Yamashita, Phys. Rev. A 108, 033307 (2023).

\bibitem{nie01} E. Nielsen, D.V. Fedorov, A.S. Jensen, and E. Garrido, Phys. Rep. 347, 373 (2001).

\bibitem{gar19b}
E. Garrido and A.S. Jensen, Phys. Rev. Research 1, 023009 (2019).

\bibitem{gar20a}
E. Garrido and A.S. Jensen, Phys. Rev. Research 2, 033261 (2020).

\bibitem{chr22}
E.R. Christensen, E. Garrido, and A. S. Jensen, Phys. Rev. A 105,033308 (2022).

\bibitem{gar22}
E. Garrido, E.R. Christensen, and A.S. Jensen, Phys. Rev. A 106, 013307 (2022).

\bibitem{gar21}
E. Garrido and A.S. Jensen Phys. Lett. A 385, 126982 (2021).

\bibitem{gar21b}
E. Garrido and A.S. Jensen, Few-body Syst. 62, 25 (2021).

\bibitem{gar23}
E. Garrido and A.S. Jensen, Eur. Phys. J. D 77, 46 (2023).

\bibitem{san18} J.H. Sandoval, F.F. Bellotti, M.T. Yamashita, T. Frederico, D.V. Fedorov, A.S. Jensen, and N.T. Zinner,
J. Phys. B: At. Mol. Opt. Phys.  51,  065004 (2018).

\bibitem{fab83} M. Fabre de la Ripelle, Ann. Phys. (N.Y.) 147, 281 (1983).


\end{thebibliography}

\end{document}